\documentclass[10pt,letterpaper]{revtex4}
\usepackage{amssymb}
\usepackage{bbold}
\usepackage{graphicx}
\usepackage{multirow}

\begin{document}
\title{Stochastic Simulator for modeling the transition to lasing
}

\author{G.P. Puccioni$^{\rm 1}$ and G.L. Lippi$^{\rm
2,3,4}$\footnote{Corresponding author}}
\affiliation{$^{\rm 1}$ \mbox{Istituto dei Sistemi Complessi, CNR, Via Madonna
del Piano 10, I-50019 Sesto Fiorentino, Italy}\\
$^{\rm 2}$ \mbox{Institut Non Lin\'eaire de Nice, Universit\'e de Nice Sophia
Antipolis}\\
$^{\rm 3}$ \mbox{CNRS, UMR 7335,
1361 Route des Lucioles, F-06560 Valbonne, France}\\
$^{\rm 4}$ email:  Gian-Luca.Lippi@inln.cnrs.fr}

\date{\today}

\begin{abstract}
A Stochastic Simulator (SS) is proposed, based on a semiclassical description
of the radiation-matter interaction, to obtain an efficient description of the
lasing transition for devices ranging from the nanolaser to the traditional
``macroscopic" laser.  Steady-state predictions obtained with the SS agree
both with more traditional laser modeling and with the description of phase
transitions in small-sized systems, and provide additional information on
fluctuations.  Dynamical information can easily be obtained, with good
computing time efficiency, which convincingly highlights the role of
fluctuations at threshold.  
\end{abstract} 

\maketitle

\section{Introduction}

The progressive miniaturization of lasers, begun with the introduction of the
VCSEL~\cite{Soda1979}, has opened a wealth of questions related to the
operation of very small devices.  Problems with the definition of threshold
were recognized very early on~\cite{Bjork1991} and still
persist~\cite{Chow2014}, and difficulties with a quantitative interpretation
of correlation functions have been more recently pointed
out~\cite{Lebreton2013}.  Failure of traditional continuous modeling based on
rate equations has been widely recognized and is discussed in detail
in~\cite{Lebreton2013} and various alternatives exist to predict the behaviour
of small lasers.  In essence, the difficulty arises from the strongly discrete
and stochastic nature of the problem and the limited number of spontaneous
emission channels in the cavity.

We can distinguish two classes of stochastic models:  1. those which describe
the evolution of the ensemble, either predicting the static features or their
statistical dynamics, and 2. those which predict the detailed dynamics.  To
the first class belong Monte-Carlo simulations, which have been recently
applied to lasers with success~\cite{Chusseau2014}, and Fokker-Planck models
which have a long history in laser physics~\cite{Risken1984}.  The Master
Equation clearly belongs to the second class, since it allows for detailed
predictions of the dynamical evolution of the system, but its use becomes
unwieldy when the number of elements (modes of the e.m. field) exceeds a few
units.  A practical reduction to a random walk on a lattice has been
proposed~\cite{Roy-Choudhury2009}.  Recently developed quantum
models~\cite{Lorke2013} are to be ascribed to this second group.\par

The aim of our Stochastic Simulator (SS) is to provide a tool for rapid
computation predicting the laser dynamics and its emergence from the
spontaneous emission, without the mathematical assumptions normally associated
to the derivation of a differential description.  Rather, we set up an
ensemble of very simple (semiclassical) rules reproducing the physics of the
interaction between radiation and matter, and let the SS predict the sequence
of events which will occur, on the basis of the randomness of each occurrence.
The advantages of this choice are that:  a. it provides extreme computational
efficiency, thus offering the possibility of quick dynamical predictions and
of performing averages over wide samples; and b. letting the operator {\it
observe emerging dynamical aspects} which otherwise would be either washed out
(in stationary approaches) or reproduce the statistical characteristics of the
modelling choice (rate equations with Langevin noise, meaningful only for
large enough averages~\cite{Haken1975}) rather than those
intrinsic to the physics of the process near threshold.  

We remark that the present proposal for a SS aims at modeling only the basic
ingredients necessary for a description of laser action.  This choice is
dictated by the interest in identifying those elements which are essentially
intrinsic to the stochastic nonlinear process.  Refinements are possible, and
planned, to include specific effects which will allow for a closer description
of particular kinds of lasers.

A discretization of the rate equations, fulfilling a similar role, has
previously been proposed~\cite{Lebreton2013}, but its scope was strongly
reduced by the neglection of the spontaneous emission -- inherent to the rate
equations~\cite{Siegman1986,Narducci1988} --, which precludes the
determination of thresholds or of the mixtures of coherent and incoherent
photons.  The absence of the latter entirely removes the interplay
spontaneous/stimulated emission, thereby preventing one from correctly
observing the rise of the stimulated component of the laser field intensity.

\section{Stochastic Simulator}

The Stochastic Simulator mimics a sequence of events as a succession of
possible occurrences taking place at discrete times, with a time
discretization ($\Delta \tau$, cf. caption of Fig.~\ref{Scurve}) small
compared to the fastest time scale.  Indeed, since each process is represented
by a probabilistic distribution determining whether and when the corresponding
process takes place, the discretization must be sufficiently fine to allow for
a correct description of even the fastest process (the escape of off-axis
spontaneous photons -- cf. below).  This ensures that all processes will be
described with sufficient accuracy, while most of the processes will have zero
outcome at any given step, at least below and around threshold.

\begin{figure}[ht!]
\hglue -.5cm
\includegraphics[width=\linewidth,clip=true]{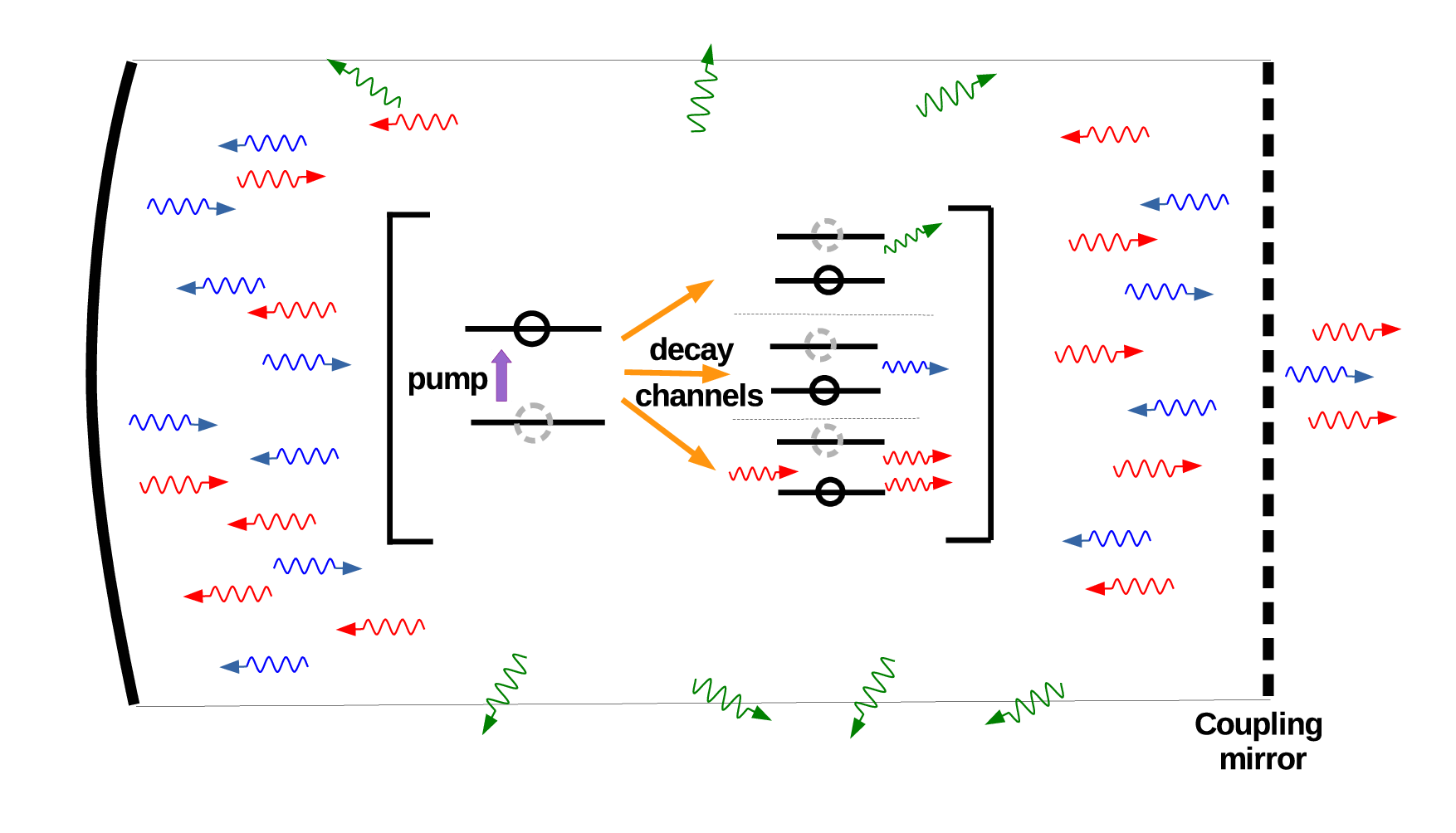}
\vglue -1cm
\caption{
Scheme of principle of the intracavity processes included in the SS:  each
{\it dipole} is pumped by an external source and may decay by emitting a
photon in one of the three following channels -- a. the {\it off-axis
spontaneous ``mode"}, grouping all modes other than the lasing one (green
photons); b. the {\it on-axis spontaneous mode} (blue photons); c. the
(on-axis) {\it stimulated mode} (red photons).  All on-axis photons are
recycled by the cavity and are transmitted by the coupling mirror; the
off-axis (green) ones exit the interaction volume laterally. All processes
(including the mirror transmission) are described stochastically.  In the
schematic representation, we summarize the ensemble of processes leading to
the population of the upper state -- with possible population inversion -- as
a single upwards magenta arrow.  The large bracket denotes the interaction
processes occurring between each individual {\it dipole} and the radiation.
Absorption is neglected as we consider an ideal system (e.g., perfect
four-level laser). Its inclusion is expected to mainly raise the laser
threshold.} 
\label{schematics}
\end{figure}

The chain of physical processes which are mimicked by the SS is the following (cf.
Fig.~\ref{schematics}):
\begin{itemize}\itemsep-1mm
\item[a.] excitation of a {\it dipole} by a pumping mechanism;
\item[b.] disexcitation of a {\it dipole} through one of the following
channels:\\[-6mm]
\begin{itemize}\itemsep-1mm
\item[i.] spontaneous relaxation into an off-axis cavity mode;
\item[ii.] spontaneous relaxation into the on-axis cavity mode;
\item[iii.] stimulated emission (into the on-axis mode).
\end{itemize}
\item[c.] escape of an off-axis spontaneous photon;
\item[d.] transmission of an on-axis spontaneous photon through the cavity
mirror;
\item[e.] transmission of a stimulated photon through the cavity mirror.
\end{itemize}
As detailed in Table~\ref{processes}, the probability distribution for the
pump is taken to be a Poissonian ($\mathbb P$), while all other processes are
simulated through a binomial distribution ($\mathbb B$), since they correspond
to yes/no events~\cite{Lebreton2013}.  All variables -- {\it dipole} number
($N$), stimulated photon number ($S$), on-axis spontaneous photon number
($R_L$), and off-axis spontaneous photon number ($R_o$) -- need to be defined
at all times, independently of their actual value (e.g., $S = 0$ far below
threshold), since the SS can only accumulate population numbers in preexisting
``categories".

\begin{table}[ht]
\caption{
Synoptic of the processes handled by the SS.  The off-axis spontaneous
emission results from the difference between the population relaxation events
($N_d$) and the on-axis spontaneous emissions ($D_L$); thus no explicit
process is associated to this channel (we assume, for simplicity, all
relaxations to be radiative).  Each event (column 3) is computed from the
probability distribution (column 4) whose first argument is defined in column
5.}

\label{processes}
{
\renewcommand{\arraystretch}{1.2}
\begin{tabular}{|c|c|c|c|c|}
\multicolumn{1}{c}{Process} & \multicolumn{1}{c}{Physical event} &
\multicolumn{1}{c}{Event} & \multicolumn{1}{c}{Math expression}
& \multicolumn{1}{c}{$\mathcal P$}\\
\hline
\hline
Excitation & Pump Absorption& $N_p$ & $\mathbb P (P)$ & $P$\\
\hline
Disexcitation & Population Relaxation& $N_d$ & $\mathbb B ({\mathcal P}_N,N)$ &
$1-e^{-\gamma_\parallel \tau_i}$\\
\hline
\hline
\multirow{2}{*}{Emission} & Spontaneous on-axis & $D_L$ & $ \mathbb
B({\mathcal P}_f,N_d)$ & $\beta$\\
\cline{2-5}
 & Stimulated & $E_S$ & $ \mathbb B({\mathcal P}_{st},N)$ &  $\gamma_\parallel \;
\beta\; S\; \Delta \tau$\\
\hline
\multirow{3}{*}{Photon Losses} & Spontaneous off-axis & $L_o$ & $ \mathbb
B({\mathcal P}_{T,o},R_o)$ & $1-e^{-\Gamma_o \tau_j}$\\
\cline{2-5}
 & Spontaneous on-axis &  $L_L$ & $ \mathbb B({\mathcal P}_{T,L},R_L)$ &
$1-e^{-\Gamma_c \tau_k}$\\
\cline{2-5}
 & Stimulated &  $L_S$ & $ \mathbb B({\mathcal P}_{T,S},S)$ &
$1-e^{-\Gamma_c\tau_m}$\\
\hline
Seed & On-axis spont. em. & $S_{sp}$ & $\left\{\begin{array}{c c c}0&{\rm
if}&\;\;\;\; S\ne 0\\
1&{\rm if}&\left\{ \begin{array}{c} S=0\\R_L\ne 0\\ \alpha > K \\
\end{array}\right.\end{array}\right.$&  \\
\hline
\end{tabular}
}
\end{table}

The following recurrence relations define the simulator's rules:
\begin{eqnarray}
\label{defN}
N_{q+1} & = & N_q + N_P - N_d - E_S \, , \\
S_{q+1} & = & S_q + E_S - L_S + S_{sp} \, , \\
R_{L,q+1} & = & R_{L,q} + D_L - L_L - S_{sp} \, , \\
\label{defRo}
R_{o,q+1} & = & R_{o,q} +(N_d - D_L) - L_o \, ,
\end{eqnarray}
where $N_P$ represents the pumping process, $N_d$ the spontaneous relaxation
processes which reduce the population inversion $N$, $E_S$ the stimulated
emission processes which also reduce $N$, $L_S$ the leakage of
stimulated photons through the output coupler, $S_{sp}$
accounts for the seed starting the first stimulated emission process, $D_L$
represents the fraction of spontaneous relaxation processes which enter the
on-axis mode (and therefore superpose to the stimulated emission), $L_L$ the
losses for the on-axis fraction of the spontaneous photons through the output
coupler, and $L_o$ the losses for the off-axis fraction of the spontaneous
photons exiting (laterally) the cavity volume.  The definitions of the
quantities appearing on the r.h.s. of eqs.~(\ref{defN}--\ref{defRo}) are given
in Table~\ref{processes}, where the parameters appearing in the last column
are the usual ones appearing in laser models~\cite{Lebreton2013}.
Specifically, $P$ represents the pump (average of the Poissonian),
$\gamma_\parallel$ is the population relaxation rate, $\Gamma_c$ the cavity
losses (on-axis mode), $\Gamma_o$ the cavity losses for the off-axis ``mode"
(i.e., the average lifetime for spontaneous photons emitted in modes other
than the on-axis one), $\beta$ represents the fraction of spontaneous emission
coupled into the on-axis mode~\cite{Bjork1991}.

At variance with the simple discretization introduced in~\cite{Lebreton2013},
the SS handles the different time scales over which the dynamical
evolution of the processes take place by introducing independent, explicit
probabilities (cf. last column in Table~\ref{processes}).  

Finally, since the SS accounts for the initiation of stimulated emission
randomly starting from a spontaneous event, we introduce the seed $S_{sp}$
(cf. Table~\ref{processes}) which activates (producing $S_{sp}=1$) only in the
absence of stimulated photons ($S =0$), in the presence of spontaneous photons
in the on-axis mode ($R_L \ne 0$), and with probability determined by a random
number $\alpha$ and a set level $K$.  The value of the constant $K$ tunes the
probability of transforming a spontaneous photon into a stimulated one, thus
initiating the amplification process.

With the values of the laser parameters defined above, the SS is started with
the desired value of the pump $P$ and initial values for the variables ($N$,
$S$, $R_L$, and $R_o$).  The choice of the initial values is not crucial as
long as they do not differ by more than a couple of orders of magnitude from
the actual averages belonging to the set of parameters chosen, since the SS is
run with a transient $\tau_t$ to let the variables relax.  The first
determination of reasonable starting values has to be empirically obtained
with sufficiently long runs (monitoring the averages).

\section{Results}

\begin{figure}[ht!]
\includegraphics[width=.45\linewidth,clip=true]{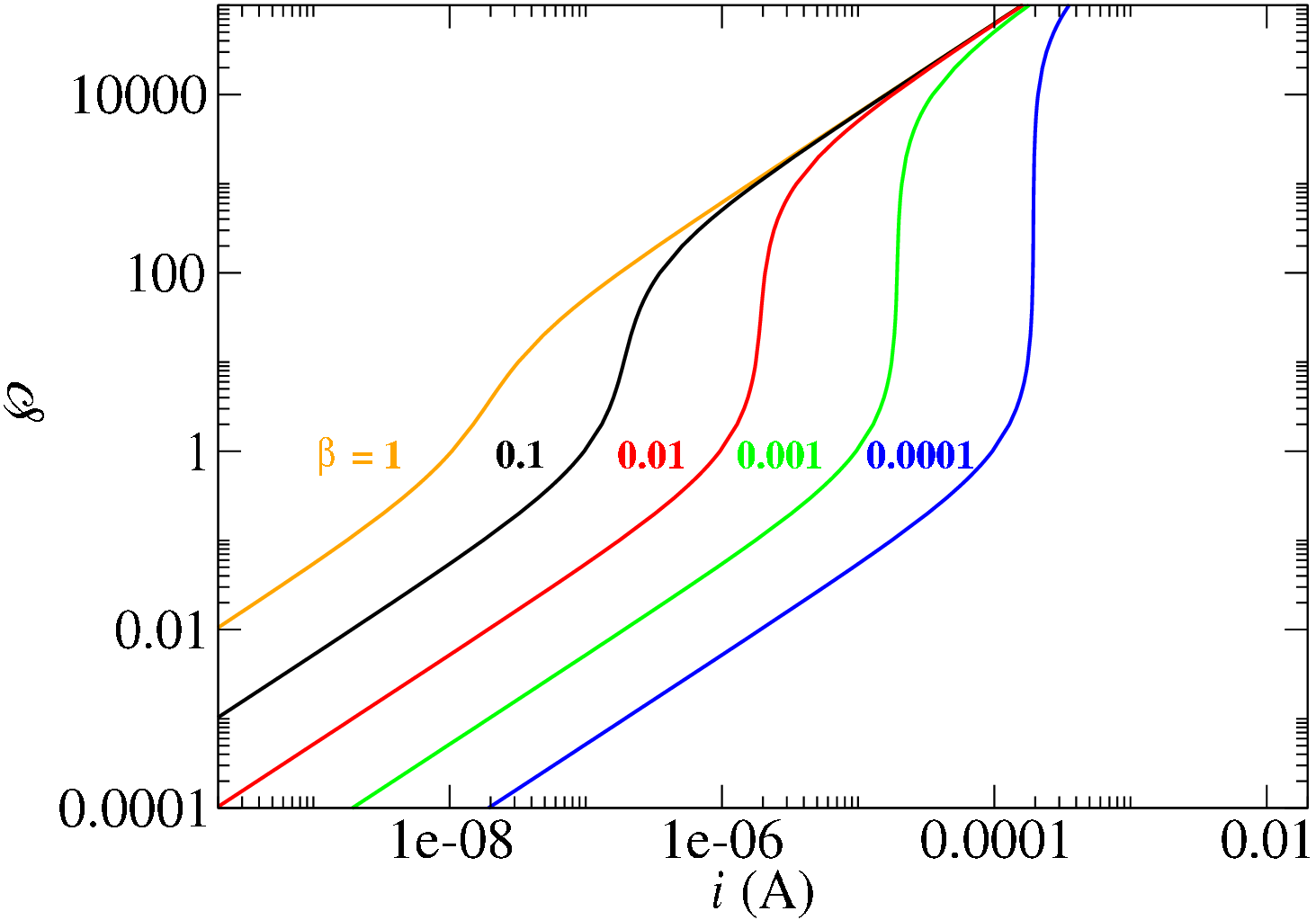}
\includegraphics[width=.5\linewidth,clip=true]{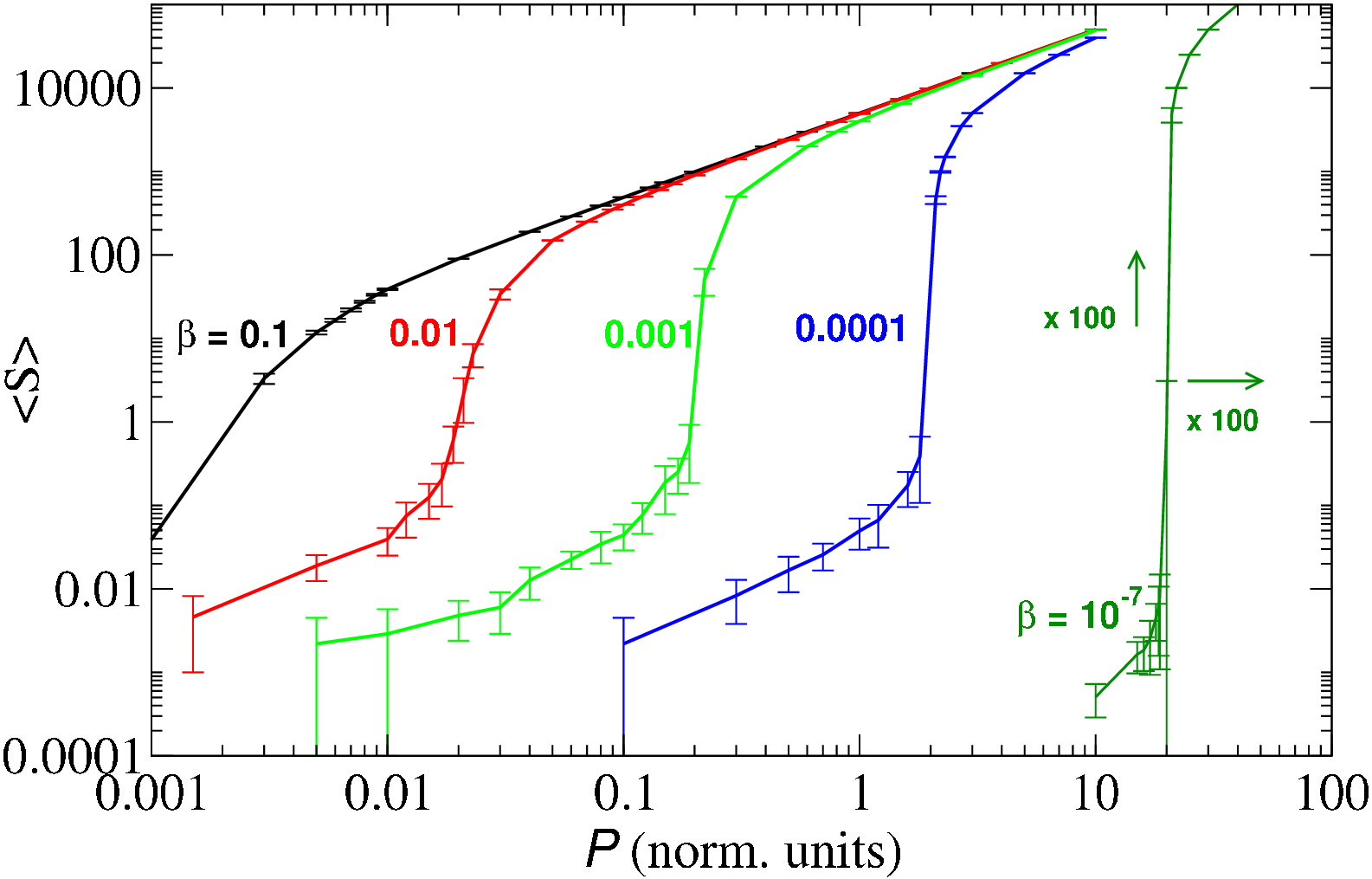}

\caption{
Left panel:  Photon number $\wp$ vs. injection current $i$ computed from the
steady-states of the rate equations (eq. (25) in \cite{Bjork1991}) with the
following parameter values:  $\gamma = 1 \times 10^{-10} s^{-1}$, $\xi = 0.1$,
$\tau_{sp} = 1 \times 10^{-9} s$, $\tau_{nr} = 1 \times 10^{-10} s$.  Right
panel:  Average photon number ($<S>$) as a function of pump ($P$) obtained
from the SS.  $<S>$ is computed as the average  of the temporal average photon
number $\overline{S} = \frac{1}{N} \sum_{\ell = 1}^N S_{\ell}$ over ten series
of events (issued from different values input into the random number
generator).  The corresponding error bars represent the standard deviation of
the individual averages $\overline{S}$ and give a measure of the variability
of the photon number, due to the stochastic nature of the conversion process.
Parameter values: $\gamma_\parallel = 2.5 \times 10^9 s^{-1}$, $\Gamma_c = 1
\times 10^{11} s^{-1}$, $\Gamma_o = 5 \times 10^{13} s^{-1}$, $K=0.05$.  The
time step used for evaluating the probabilities (last column of
Table~\ref{processes}) is $\Delta \tau = \frac{1}{10 \times \Gamma_o}= 2 fs$.
Notice that the curve with $\beta = 10^{-7}$ is rescaled by a factor $10^{-2}$
both in the horizontal and vertical axes for graphical purposes.  }
\label{Scurve} \end{figure}

Fig.~\ref{Scurve} shows the predicted laser output as a function of the pump
(right panel) which gives a good qualitative agreement with the equivalent
curves computed from the rate equations (cf. caption for details).  The SS is
capable of not only producing a meaningful laser response, but also predicting
the deviations (error bars) for nanolasers ($0.01 \lesssim \beta \le 1$), for
microlasers ($0.0001 \lesssim \beta \lesssim 0.01$) and even for macroscopic
lasers (e.g., $\beta = 10^{-7}$).  The progressive sharpening of the
transition confirms the well-known trend with system size and agrees with the
prediction that in the thermodynamic limit the laser threshold becomes a true
phase transition~\cite{Dohm1972}.  Thus, the SS can be used to model class B
lasers~\cite{Tredicce1985} of any size. 

\begin{figure}[ht!]
\includegraphics[width=.32\linewidth,clip=true]{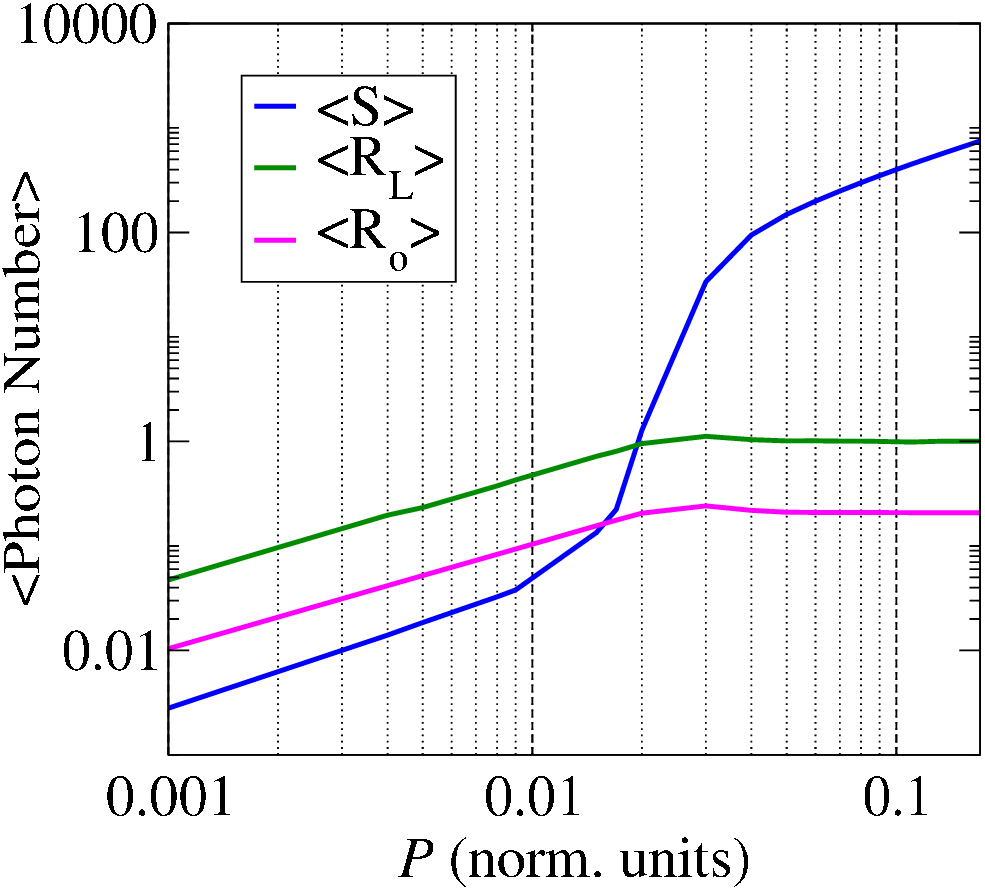}
\includegraphics[width=.32\linewidth,clip=true]{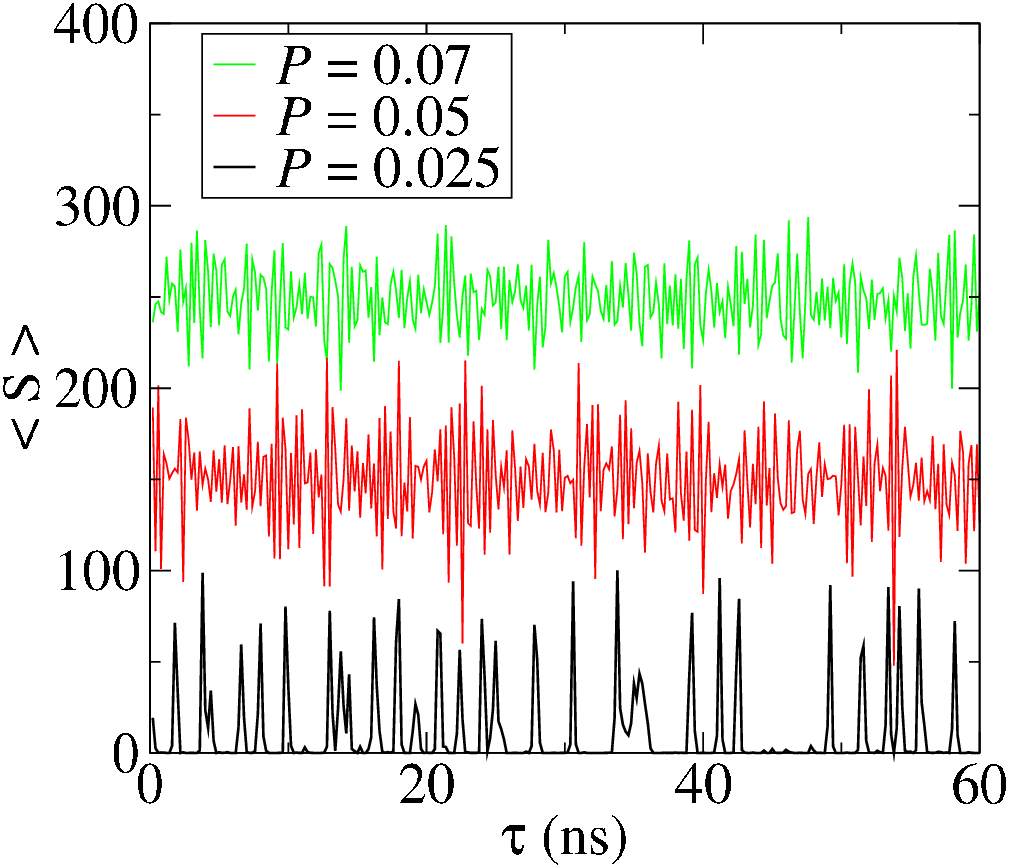}
\includegraphics[width=.32\linewidth,clip=true]{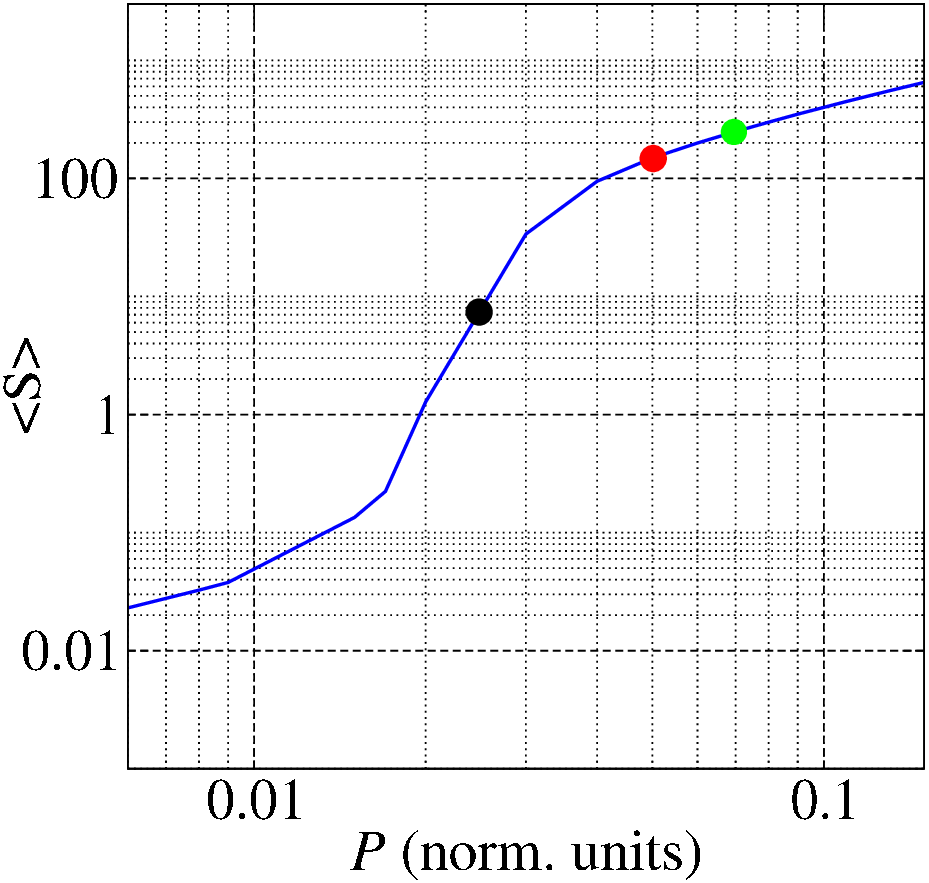}
\caption{
Static and dynamical predictions for $\beta = 0.01$.  Left panel:  Average
photon number for stimulated photons (blue curve), on-axis (green) and
off-axis (magenta) spontaneous photons as a function of pump.  Center panel:
dynamical evolution of $<S>$ for three different values of pump, marked by the
corresponding colors on the steady-state response (right panel).  }
\label{3figs} \end{figure}

Fig.~\ref{3figs} offers other predictive aspects of the SS.  The left panel
shows the average photon number, as a function of pump, for the three
different kinds of photons:  spontaneous off-axis, spontaneous on-axis and
stimulated.  According to  ``conventional wisdom", which defines threshold as
the pump value for which the average number of stimulated photons equals the
average sum of all the spontaneous photons, one can read off the graph the
threshold value as $P \approx 0.02$ for the parameters of this simulation. 
It is important to remark that while the stimulated component (blue curve)
steadily grows, both spontaneous components remain clamped above threshold.
The residual growth -- which then decays -- of both spontaneous emission
components in the ``S"-portion of the stimulated emission remains for the
moment unexplained.

The center panel shows the time-dependent stimulated emission output by the
laser for different values of pump (averaged over a $\Delta t = 0.1 ns$ time
window, to simulate the finite bandwidth of a typical detector).  This
convincingly shows that in the threshold region (cf. corresponding color-coded
points in the right panel) the fluctuations are as large as the average (black
and red curves), thus confirming the observations of~\cite{Lebreton2013}.

As a final point, the running time of the SS is remarkably short.  Running the
SS (programmed in C using GSL routines~\cite{GSL}) on a desktop PC AMD Phenom X6 1090T, and defining an efficiency $\eta =
\frac{\rm physical\; simulation\; time}{\rm computer\; running\; time}$, we
find $0.1 \frac{ns}{s} \lesssim \eta \lesssim 0.25 \frac{ns}{s}$ depending on
the value of $\beta$ and on the average values of the variables (the better
efficiencies being attained for large $\beta$-values and low variable
averages).  Thus, we can predict, with $fs$ accuracy, the equivalent of the
trace displayed on an oscilloscope (typically $50ns$ long for good resolution)
in as little as $20s$ of computation!

\section{Conclusions}

In conclusion, the SS offers the possibility of obtaining very fast
predictions on the dynamics of stimulated and spontaneous photons (both on-
and off-axis) in lasers whose size ranges from the nanoscale, all the way to
the conventional macroscopic devices.  Thanks to the computational speed,
average predictions are easily obtained and agree with theoretical
considerations about phase transitions in small-size systems.  As such, we can
consider the SS as a valuable tool for providing complementary information to
the one usually obtained from traditional modeling of lasers.

\end{document}